\documentclass[twocolumn,showpacs,preprintnumbers,amsmath,amssymb,aps]{revtex4}

\usepackage{amsfonts}                                              
\usepackage[dvips]{graphicx}                                       
\usepackage[ps2pdf,bookmarks=true,bookmarksnumbered=true,breaklinks=true,pdfborder={0 0 112.0}]{hyperref} 
\usepackage{color}
\usepackage{bm}

\newcommand{\matrixsymb}[1]{\mathsf{#1}}                           
\newcommand{\eqnref}[1]{Eq.~\eqref{#1}}                            
\newcommand{\figref}[1]{Fig.~\ref{#1}}                             
\newcommand{\secref}[1]{Sec.~\ref{#1}}                             
\newcommand{\appref}[1]{App.~\ref{#1}}                             
\newcommand{\blankfrac}[2]{\substack{{#1}\\{#2}}}
\newcommand{\ket}[1]{|#1\rangle}                                   
\newcommand{\e}[1]{\text{e}^{#1}}                                  
\newcommand{\cmplxi}{\text{i}}                                     
\newcommand{\bracetextsize}{\displaystyle}                         
\newcommand{\tr}{\operatorname{Tr}}                                
\renewcommand{\vec}[1]{\mathbf{#1}}                                
 
\newcommand{\obrace}[2]{\overbrace{#1}^{\bracetextsize{#2}}}       
\newcommand{\punc}[1]{\,#1}

\begin{document}
\title{Itinerant ferromagnetism in an atomic Fermi gas: Influence of 
population imbalance}
\author{G.J. Conduit}
\email{gjc29@cam.ac.uk}
\author{B.D. Simons}
\affiliation{Theory of Condensed Matter Group, Department of Physics, 
Cavendish Laboratory, 19, J.J. Thomson Avenue, Cambridge, CB3 0HE. UK}
\date{\today}

\begin{abstract}
We investigate ferromagnetic ordering in an itinerant ultracold atomic Fermi gas
with repulsive interactions and population imbalance. In a spatially uniform
system, we show that at zero temperature the transition to the itinerant
magnetic phase transforms from first to second order with increasing population
imbalance. Drawing on these results, we elucidate the phases present in a
trapped geometry, finding three characteristic types of behavior with changing
population imbalance. Finally, we outline the potential experimental
implications of the findings.
\end{abstract}


\maketitle

\section{Introduction}\label{sec:Introduction}

Feshbach resonance phenomena provide unprecedented control of pair interactions
in degenerate atomic Fermi gases~\cite{76s12,93tvs05}. This feature has allowed
extensive studies of pairing phenomena in two-component Fermi gases providing
access to the crossover between a Bose-Einstein condensate (BEC) of molecules
and the Bardeen-Cooper-Schrieffer (BCS) state of Cooper
pairs~\cite{03sph08,03ghzsdskvk06,04cbarjhg08,04rgj01}. Although the emphasis of
experimental investigations has been primarily on the problem of resonance
superfluidity, interacting Fermi gases support other strongly-correlated phases
including itinerant ferromagnetism.

In solid state condensed matter systems, the problem of itinerant ferromagnetism
has a long history dating back to the pioneering studies by Stoner \cite{57s04}
and Wohlfarth \cite{62wr11}. These early investigations proposed that, at low
enough temperatures, a Fermi gas subject to a repulsive interaction potential
could undergo a continuous phase transition into an itinerant spin polarized
phase~\cite{79z11}. This Stoner transition reflects the shifting balance between
the potential energy gained in spin polarization through Pauli exclusion
statistics, and the associated cost in kinetic energy.  Subsequent studies
showed that fluctuations in the magnetization at low temperatures drive the
second order transition first order at low enough
temperatures~\cite{64s09,97bkv04,99bkv06,00v06,02bk12,05bkr06}. Such behavior is
born out around quantum criticality in a variety of experimental solid state
systems including $\text{ZrZn}_{2}$~\cite{04uph12,05uph04},
$\text{UGe}_{2}$~\cite{00hsb07},
MnSi~\cite{01pjl11,04yztstfb02,07uggcrsamrldmapbybflss01,09orsar02,09pkls03},
$\text{CoS}_{2}$~\cite{08orgppbr06},
$\text{YbRh}_{2}\text{Si}_{2}$~\cite{08myi09}, and
$\text{SrRuO}_{3}$~\cite{07hdsjkccn07}. When subject to a magnetic field, the
attendant increase in Zeeman energy results in the bifurcation of the
tricritical point separating the region of first and second order ferromagnetic
transitions into two lines of metamagnetic critical points.

In the following, we will explore the potential implications of this itinerant
magnetic phase behavior on the equilibrium properties of strongly interacting
two-component atomic Fermi gases; here we refer to the pseudo-spin associated
with the hyperfine states characterizing the two atomic populations. However, in
contrast to the solid state system, the application of these ideas to the atomic
Fermi gas must address the features imposed by the trap geometry, and the
constraints resulting from the inability of particles to transfer between
different spin states~\cite{06lh07}. As a result, in the general case, one must
consider atomic Fermi mixtures in which an effective spin polarization is
imposed by population imbalance~\cite{07c02,07sr08}. The potential for itinerant
ferromagnetism in atomic Fermi gases has been already addressed in the
literature, \citet{08zhw05} showed that an optical lattice would reduce the
repulsive interaction strength required to realize ferromagnetism, and
\citet{02sy10} studied a trapped system in the Thomas-Fermi
approximation. Subsequently, \citet{05dm11} developed a diagrammatic
perturbative expansion in interaction strength to address the phase behavior of
the balanced two-component Fermi system. In the following, we will develop a
functional integral formulation to explore the phase behavior of the general
population imbalanced system. As well as providing access to the mean-field
phase behavior of the system, such an approach allows for future considerations
of the collective low energy spin dynamics of the spin polarized phase.
Moreover, the theory provides a platform to explore the potential for the
development of an equilibrium spin textured phase recently conjectured in
relation to the solid state
system~\cite{97bkv04,01hsrkbcf03,02wm06,05kiptsm09,08bcs05}.

The paper is organized as follows: In \secref{sec:ItinerantAtomFormalism} we
derive an expression for the thermodynamic potential of the system as a function
of the local density and in-plane magnetization fields. To address the important
effects of spin-wave fluctuations on the nature of the equilibrium phase
diagram, we will explore the renormalization of the mean-field equations keeping
those terms that are second order in the coupling strength, $g$. Using this
result, in \secref{sec:TrappingPotentialUniformSystem} we analyze the phase
diagram of the spatially uniform system as a function of the interaction
strength, $g$, and chemical potential shift. Finally, in
\secref{sec:TrappingPotentialTrappedSystem} we explore in detail the phase
behavior of the magnetic system in the atomic trap geometry.

\section{Field integral formulation}\label{sec:ItinerantAtomFormalism}

Expressed as a coherent state path integral, the quantum partition function of
a population imbalanced two-component Fermi gas is given by
\begin{eqnarray}
 \mathcal{Z}&=&\int\!\!\mathcal{D}\psi 
 \exp\Biggl[-\int_0^\beta\!\! d\tau\,d\vec{r}\!\!\!\!\sum_{\sigma=\{\uparrow,\downarrow\}}
 \!\!\!\!\bar{\psi}_\sigma(-\cmplxi\partial_{\tau}+\hat{\xi}-\sigma\Delta\mu)\psi_{\sigma}
 \nonumber\\
 &-&\int_0^\beta d\tau\,d\vec{r}\,g
 \bar{\psi}_{\uparrow}\bar{\psi}_{\downarrow}\psi_{\downarrow}
 \psi_{\uparrow}\Biggr]\,.
\end{eqnarray}
where $\bar{\psi}_\sigma(\tau,\vec{r})$ and $\psi_\sigma(\tau,\vec{r})$ denote
Grassmann fields, $\beta=1/k_{\rm B}T$ is the inverse temperature, and
$\hat{\xi}=\hat{p}^{2}/2m-\mu$. Here we have used a pseudo-spin index,
$\sigma\in\{\uparrow,\downarrow\}$, to discriminate the two components. As
independent particles (with no interconversion), the density of the two
majority/minority degrees of freedom must be specified by two chemical
potentials. For convenience, it is helpful to separate the chemical potentials
into their sum and difference; $\mu+\Delta\mu$ for up-spin and $\mu-\Delta\mu$
for down-spin. In this representation, population imbalance may be adjusted
through the chemical potential shift, $\Delta\mu$. Note that, although
population imbalance is synonymous with a global pseudo-spin magnetization, a
spontaneous symmetry breaking into an itinerant ferromagnetic phase can still
develop with the appearance of a non-zero in-plane component of the
magnetization. Finally, we suppose that the strength of the repulsive $s$-wave
contact interaction, $g\delta^{3}(\vec{r})$, can be tuned using a Feshbach
resonance.

\subsection{Hubbard-Stratonovich decoupling}\label{sec:DecouplingTheContactInteraction}

To develop an effective low-energy theory for the Fermi gas, it is convenient to
decouple the quartic contact interaction by introducing auxiliary bosonic
fields, $\rho$ and ${\bm\phi}$, conjugate to the local density
$\sum_{\alpha=\{\uparrow,\downarrow\}}\bar{\psi}_{\alpha}\psi_{\alpha}$ and
magnetization
$\sum_{\alpha,\beta=\{\uparrow,\downarrow\}}\bar{\psi}_{\alpha}{\bm\sigma}_{\alpha\beta}\psi_{\beta}$
respectively, setting
\begin{eqnarray}
 &\mathcal{Z}&=\int\mathcal{D}{\bm\phi}\mathcal{D}\rho\mathcal{D}\psi
 \exp\Biggl\{-\int d\tau d\vec{r}\,\Biggl[g({\bm\phi}^{2}-\rho^{2})\nonumber\\
 &+&\!\!\!\!\!\!\!\!\sum_{\alpha,\beta=\{\uparrow,\downarrow\}}\!\!\!\!\!\!\bar{\psi}_{\alpha}
\left[(\hat{G}_{0}^{-1}+g\rho)\delta_{\alpha\beta}-
 (\Delta\mu\vec{e}_{\text{z}}+g{\bm\phi})\cdot
 {\bm\sigma}_{\alpha\beta}\right]\psi_{\beta}\Biggr]\!\Biggr\}.\nonumber\\
 \label{eqn:coldatomferroqptprioordiagonalisation1}
\end{eqnarray}
Here $\hat{G}_{0}=(-\cmplxi\partial_{t}+\hat{\xi})^{-1}$ defines the Green's
function of the non-interacting system, and ${\bm\sigma}$ denotes the
vector of Pauli spin matrices. Note that, without decoupling in both the Hartree
and Fock channels, one would subsequently encounter unphysical diagrammatic
contributions to the perturbative scheme developed
below~\cite{70k11,70esw03,74mgnr11}. It is also the simplest approach that
maintains spin rotational invariance of the Hamiltonian, and leads to the
correct set of Hartree-Fock equations~\cite{79h03,79pk05IV}. Then, integrating
over the Fermi fields, one obtains the expression
\begin{eqnarray}
 &&\mathcal{Z}=\int \mathcal{D}{\bm\phi}\mathcal{D}\rho\mathcal{D}\psi 
 \,\e{{-\int d\tau d\vec{r}\, g({\bm\phi}^{2}-\rho^{2})}}\nonumber\\
 && \times \exp\left[\tr\ln\left(\hat{G}_{0}^{-1}+g\rho-
 {\bm\sigma}\cdot\left(\Delta\mu\vec{e}_{\text{z}}+g{\bm\phi}
 \right)\right)\right]\,.
 \label{eqn:coldatomferroqptprioordiagonalisation}
\end{eqnarray}
At this stage the analysis is exact, but to proceed further one must employ an
approximation. To orient our discussion and make contact with conventional
Stoner theory, let us first consider a direct saddle-point approximation scheme.

\subsection{Stoner mean-field theory}\label{sec:FerroOrderingAuxiliaryFieldMeanFieldSaddlePointValues}

As well as the ``effective'' magnetization imposed by population imbalance, we
anticipate the development of a spontaneous magnetization which will drive the
axis of quantization away from the z-axis. We re-orient the axis of quantization
to lie parallel to the net magnetization, denoted in mean-field theory (with
over-bars)
$\overline{{\bm\phi}}=\overline{{\bm\phi}}_{\perp}+\overline{\phi}_{\text{z}}\vec{e}_{\text{z}}$,
$\overline{{\bm\phi}}_{\perp}=(\overline{\phi}_{\text{x}},\overline{\phi}_{\text{y}})$,
and with this definition, the total magnetization of the system is given by
$\overline{\vec{M}}=\Delta\mu\vec{e}_{\text{z}}/g+\overline{{\bm\phi}}$.
Separately varying the action with respect to
$\overline{{\bm\phi}}_{\perp}$ and $\overline{\phi}_{\text{z}}$ one
obtains, respectively, the saddle-point equations,
\begin{equation*}
\left(\begin{array}{c}\overline{{\bm\phi}}_{\perp}\\\overline{\phi}_{\text{z}}\end{array}\right)=-\frac{(\beta V)^{-1}\tr(\hat{G}_{+}-\hat{G}_{-})}{\sqrt{(g\overline{{\bm\phi}}_{\perp})^{2}+(g\overline{\phi}_{\text{z}}+\Delta\mu)^{2}}}\left(\begin{array}{c}g\overline{{\bm\phi}}_{\perp}\\g\overline{\phi}_{\text{z}}+\Delta\mu\end{array}\right)\,,
\end{equation*}
where
$\hat{G}_\pm^{-1}=\hat{G}_{0}^{-1}+g\overline{\rho}\mp\left|\Delta\mu\vec{e}_{\text{z}}+
g\overline{{\bm\phi}}\right|$, and $V$ denotes the total volume of the
system. Together, these equations admit two possible solutions:

\begin{description}

 \item[{[$\overline{{\bm\phi}}_{\perp}=\vec{0}$ and
 $\overline{M}=\overline{\phi}_{\text{\textnormal{z}}}$]:}] The total magnetization of the
 system can be ascribed to population imbalance with no spontaneous
 magnetization in-plane. Within this solution, $\overline{M}$ is a function of
 $|g\overline{\phi}_{\text{z}}+\Delta\mu|$, so it can be used to infer the
 chemical potential shift, $\Delta\mu$.

 \item[{[$\overline{{\bm\phi}}_{\perp}\ne\vec{0}$]:}] The total magnetization
 takes the form
 $\overline{M}=(\overline{{\bm\phi}}_{\perp}^{2}+\overline{\phi}_{\text{z}}^{2})^{1/2}$. Along
 z-axis, the magnetization is fixed due to population imbalance, with the
 additional magnetization developing within the x-y plane. In this case, the
 saddle-point solution translates to the condition $\Delta\mu=0$, i.e.  no
 chemical potential shift is required to recover the fixed z-component of the
 magnetization due to the population imbalance; it is simply given by the
 resolved component of the total magnetization.

\end{description}

\noindent
The total population $N=N_{\uparrow}+N_{\downarrow}$ can in turn be obtained
from the variation $\delta\overline{S}/\delta\overline{\rho}=0$.

Expanding the action in interaction strength, $g$,
$\overline{S}=g\phi_{\text{z}}^{2}\tr(1+g\hat{G}_{0}\hat{G}_{0})=g\phi_{\text{z}}^{2}(1-g\nu)$,
and one can extract the familiar Stoner criterion \cite{63k09,63h11} for a
population balanced system, with $\nu$ being the density of states. For $g\nu<1$
the state is unmagnetized, $\overline{M}=0$, and chemical potentials of the two
Fermi surfaces remain equal. If $g\nu>1$ then the state is magnetized with
$\overline{M}=\sqrt{(g\nu-1)/g^{3}\nu''}$. We also note that the Stoner
criterion can be reformulated to account for population imbalance giving
$\overline{S}=g\phi^{2}(1-g\nu)-g^{2}\Delta\mu^{2}$, leading to a transition at
the same value of interaction strength as for the balanced system. Although, at
this order, the saddle-point approximation predicts a continuous transition to a
ferromagnetic phase for the balanced system, it is well-established that
fluctuations of the magnetization field drive the transition first order at low
temperature \cite{58ak05}. This effect can be captured by retaining fluctuation
contributions to second order in the interaction. In the following, we will
explore the impact of fluctuations on the equations of motion associated with
the uniform mean-field.

\subsection{Integrating out auxiliary field fluctuations}\label{sec:IntegratingOutAuxiliaryFieldFluctuations}

To implement this program, it is convenient to parameterize the
Hubbard-Stratonovich fields into some, as yet undetermined, stationary
(spatially uniform) values ${\bm\phi}_{0}$ and $\rho_{0}$, and
fluctuations around them, ${\bm\phi}_{\text{fl}}$ and
$\rho_{\text{fl}}$. Integrating out these fluctuations, the goal is to obtain
the renormalized mean-field equations for ${\bm\phi}_{0}$ and $\rho_{0}$
retaining contributions to second order in $g$. Substituting
${\bm\phi}={\bm\phi}_{0}+{\bm\phi}_{\text{fl}}$ and
$\rho=\rho_{0}+\rho_{\text{fl}}$ into
\eqnref{eqn:coldatomferroqptprioordiagonalisation}, and rotating the z-axis from
the quantization direction to lie along the direction of uniform
magnetization using the constant matrix $\matrixsymb{T}$, one obtains
\begin{eqnarray*}
 &\mathcal{Z}&\!\!=\e{{-\beta Vg({\bm\phi}_{0}^{2}-\rho_{0}^{2})}}\!\!
 \int\!\!\mathcal{D}\rho_{\text{fl}}\mathcal{D}{\bm\phi}_{\text{fl}}\,
 \exp{\left[-\!\!\int\!\! d\tau d\vec{r}\,g({\bm\phi}_{\text{fl}}^{2}-\rho_{\text{fl}}^{2})\right]}
 \nonumber\\
 &\times&\!\!\exp\left[\tr\ln \matrixsymb{G}^{-1}+\tr\ln(\matrixsymb{I}
 +g\matrixsymb{G}\matrixsymb{T}^{-1}(\matrixsymb{I}\rho_{\text{fl}}+
 {\bm\sigma}\cdot{\bm\phi}_{\text{fl}})\matrixsymb{T})\right]\punc{,}
\end{eqnarray*}
where now $\hat{G}_{\pm}^{-1}=\hat{G}_{0}^{-1}+g\rho_{0}\mp|\Delta\mu
\vec{e}_{\text{z}}+g{\bm\phi}_{0}|$ denotes the elements of the inverse
Green's function of the system at the level of the renormalized mean-field,
$\matrixsymb{\hat{G}}={\rm diag}(\hat{G}_{+},\hat{G}_{-})$. Then, expanding the
action to second order in fluctuations, $\rho_{\text{fl}}(\vec{r},\tau)$ and
${\bm\phi}_{\text{fl}}(\vec{r},\tau)$, and performing the functional
integral, one obtains the thermodynamic grand potential from the quantum
partition function using $\Phi_{\text{G}}=-\beta^{-1}\ln\mathcal{Z}$,
\begin{eqnarray}
 \label{eqn:eqn:coldatomferroqptintegratingoutfluctuations}
&&\Phi_{\text{G}}=\obrace{\tr\ln \hat{G}_{+}^{-1}+\tr\ln \hat{G}_{-}^{-1}}{\dagger}+g\left(\phi_{0}^{2}-\rho_{0}^{2}\right)\nonumber\\
&&+ \obrace{\frac{1}{2}\tr\ln\left(1-g^{2}\Pi_{++}\Pi_{--}\right)}{\parallel}\nonumber\\
&&+\obrace{\frac{1}{2}\tr\ln\left(1+g\Pi_{+-}+g\Pi_{-+}+g^{2}\Pi_{+-}\Pi_{-+}\right)}{\perp}\punc{,}
\end{eqnarray}
a result that is independent of the transformation $\matrixsymb{T}$. Here we
have defined the spin-dependent polarization operator,
\begin{eqnarray*}
 \Pi_{ss'}(\omega,\vec{q})=\frac{2}{\beta V}\sum_{\omega',\vec{k}}
 G_{s}(\omega',\vec{k})G_{s'}(\omega'-\omega,\vec{k}-\vec{q})\punc{,}
\end{eqnarray*}
where the sum on $\omega'$ runs over fermionic Matsubara frequencies. The term
labeled ($\dagger$) simply represents the thermodynamic potential of a
non-interacting Fermi gas with shifted chemical potentials. The term labeled
($\perp$) is due to transverse fluctuations of the magnetization field and
coincides with that obtained in Ref.~\cite{79pk05V}. By contrast, the term
labeled ($\parallel$), corresponding to longitudinal fluctuations, differs from
that obtained in Ref.~\cite{79pk05V} by the additional contributions from
density fluctuation effects.

To proceed, we now expand the potential $\Phi_{\text{G}}$ to second order in $g$ and perform 
the summations over Matsubara frequencies. Rearranging the momenta summations, 
one obtains
\begin{eqnarray}
 &\Phi_{\text{G}}&=-\frac{1}{\beta V}\sum_{\blankfrac{\vec{k}}{s=\{+,-\}}}\ln\left(1+\e{-\beta(\epsilon_{\vec{k}}-\mu_{s})}\right)\nonumber\\
 &+&g\left({\bm\phi}_{0}^{2}-\rho_{0}^{2}\right)+gN_{+}N_{-}\nonumber\\
 &+&\!\!\frac{2g^{2}}{V}\!\!\sum_{\vec{k}_{1,2,3}}\!\!\frac{\obrace{n_{+}(\epsilon_{\vec{k}_{1}})n_{-}(\epsilon_{\vec{k}_{2}})}{\Diamond}(1\!-\!n_{+}(\epsilon_{\vec{k}_{3}}))(1\!-\!n_{-}(\epsilon_{\vec{k}_{4}}))}{\epsilon_{\vec{k}_{1}}+\epsilon_{\vec{k}_{2}}-\epsilon_{\vec{k}_{3}}-\epsilon_{\vec{k}_{4}}}\punc{,}\nonumber\\
\end{eqnarray}
where $\mu_{s}=\mu-g\rho_{0}+s|\Delta\mu
\vec{e}_{\text{z}}+g{\bm\phi}_{0}|$,
$n_{s}(\epsilon)=(1+\exp(-\beta(\epsilon-\mu_{s})))^{-1}$, and
$N_{s}=\sum_{\vec{k}}n_{s}(\epsilon_{\vec{k}})$. Conservation of momentum
requires that $\vec{k}_{1}+\vec{k}_{2}=\vec{k}_{3}+\vec{k}_{4}$. Physically, the
numerator of the second order term indicates that the matrix element associated
with the transition $(\vec{k}_{1},\vec{k}_{2})\rightarrow
(\vec{k}_{3},\vec{k}_{4})$ is proportional to the probability that states
$\vec{k}_{1}$ and $\vec{k}_{2}$ are occupied, whilst states $\vec{k}_{3}$ and
$\vec{k}_{4}$ are unoccupied. Following \citet{96p07} (and the earlier
discussion of \citet{58ak05}), to renormalize the unphysical divergence of the
term in $n^{2}(\epsilon)$, labeled ($\Diamond$) close to resonance, we
regularize the effective interaction at second order in scattering length $a$,
\begin{equation*}
 g(\vec{k}_{1},\vec{k}_{2})\mapsto\frac{2k_{\text{F}}a}{\pi\nu}-\frac{8k_{\text{F}}^{2}a^{2}}{\pi^{2}\nu^{2}V^{2}}\sum_{\vec{k}_{3,4}}\frac{1}{\epsilon_{\vec{k}_{1}}+\epsilon_{\vec{k}_{2}}-\epsilon_{\vec{k}_{3}}-\epsilon_{\vec{k}_{4}}}\punc{,}
\end{equation*}
where $\nu=\sqrt{\mu}/\sqrt{2}\pi^{2}$ and $k_{\text{F}}=\sqrt{2m\mu}$. In a
population imbalanced system the definition for the chemical potential is that
which gives the same total number of particles in the population balanced
system, that is $k_{\text{F}}=\sqrt[3]{3\pi^{2}(n_{\uparrow}+n_{\downarrow})}$,
where $n_{\uparrow}$ and $n_{\downarrow}$ are the number of up and down-spin
particles; this definition holds true in both the canonical and grand canonical
ensembles. This regularization of the contact interaction exactly cancels the
divergent terms in $n^{2}(\epsilon)$, labeled ($\Diamond$). Furthermore, the
terms in $n^{4}(\epsilon)$ are zero by symmetry.  Finally, making use of the
symmetry in $\vec{k}_{3}$ and $\vec{k}_{4}$, one obtains
\begin{eqnarray}
 \label{eqn:popimbthermpot}
 &\Phi_{\text{G}}&=
 -\frac{1}{\beta V}\sum_{\blankfrac{\vec{k}}{s=\{+,-\}}}\ln\left(1+\e{-\beta(\epsilon_{\vec{k}}-\mu_{s})}\right)\nonumber\\
 &+&\!\!\!\!\frac{2k_{\text{F}}a}{\pi\nu}\left({\bm\phi}_{0}^{2}-\rho_{0}^{2}\right)
 +\frac{2k_{\text{F}}a}{\pi\nu}N_{+}N_{-}
\nonumber\\
 &-&\!\!\!\!\frac{8k_{\text{F}}^{2}a^{2}}{\pi^{2}\nu^{2}V^{3}}\sum_{\vec{k}_{1,2,3}}\frac{n_{+}(\epsilon_{\vec{k}_{1}})n_{-}(\epsilon_{\vec{k}_{2}})\left(n_{+}(\epsilon_{\vec{k}_{3}})+n_{-}(\epsilon_{\vec{k}_{3}})\right)}{\epsilon_{\vec{k}_{1}}+\epsilon_{\vec{k}_{2}}-\epsilon_{\vec{k}_{3}}-\epsilon_{\vec{k}_{4}}}\punc{.}\nonumber\\
\end{eqnarray}
From the thermodynamic potential we can compute the free energy per unit volume
$F=\Phi_{\text{G}}+\sum_{\sigma=\{\uparrow,\downarrow\}}(\mu+\sigma\Delta\mu)N_{\sigma}$.
To consolidate terms entering the free energy we switch from the population
imbalance pseudo-spin basis to the magnetization basis, retain contributions to
order ${\cal O}((k_{\text{F}}a)^2)$, recall that if $\Delta\mu=0$ then $M\ne0$, whereas if
$\Delta\mu\ne0$ then $M=0$, and affect the rearrangement
\begin{widetext}
\begin{eqnarray*}
&&\frac{2k_{\text{F}}a}{\pi\nu}\left({\bm\phi}_{0}^{2}-\rho_{0}^{2}\right)+\sum_{\sigma=\{\uparrow,\downarrow\}}(\mu+\sigma\Delta\mu)N_\sigma\nonumber\\
&&=\obrace{\left(\mu-\frac{2k_{\text{F}}a}{\pi\nu}\rho_{0}+\left|\Delta\mu\vec{e}_{\text{z}}+\frac{2k_{\text{F}}a}{\pi\nu}{\bm\phi}_{0}\right|\right)}{\mu_{+}}
N_{+}+\obrace{\left(\mu-\frac{2k_{\text{F}}a}{\pi\nu}\rho_{0}-\left|\Delta\mu\vec{e}_{\text{z}}+\frac{2k_{\text{F}}a}{\pi\nu}{\bm\phi}_{0}
\right|\right)}{\mu_{-}}N_{-}\nonumber\\
&&\left.{\begin{array}{l}+\frac{2k_{\text{F}}a}{\pi\nu}\left({\bm\phi}_{0}^{2}-\rho_{0}^{2}\right)+\left(\frac{2k_{\text{F}}a}{\pi\nu}\rho_{0}+\Delta\mu-\left|\Delta\mu\vec{e}_{\text{z}}+\frac{2k_{\text{F}}a}{\pi\nu}{\bm\phi}_{0}\right|\right)N_{+}\\
+\left(\frac{2k_{\text{F}}a}{\pi\nu}\rho_{0}-\Delta\mu+\left|\Delta\mu\vec{e}_{\text{z}}+\frac{2k_{\text{F}}a}{\pi\nu}{\bm\phi}_{0}\right|\right)N_{-}\end{array}}\right\}{\varnothing}\punc{.}
\end{eqnarray*}
\end{widetext}
Then, if we set ${\bm\phi}_{0}=\overline{{\bm\phi}}+
\Delta{\bm\phi}$ and $\rho_{0}=\overline{\rho}+\Delta\rho$, an expansion
in $\Delta{\bm\phi}$ and $\Delta\rho$ shows that the terms labeled
($\varnothing$) sum to zero to the accuracy of the free energy,
$\mathcal{O}((k_{\text{F}}a)^2)$. Retaining the remaining contribution, the free energy
reduces to the form,
\begin{eqnarray*}
 &\!\!\!\!F&\!\!\!=\!\obrace{-\frac{1}{\beta V}\!\!\!\!\!\!\sum_{\blankfrac{\vec{k}}{s=\{+,-\}}}\!\!\!\!\!\!\ln\left(\!1\!+\!\e{-\beta(\epsilon_{\vec{k}}-\mu_{s})}\!\right)\!+\!\!\!\!\!\!\sum_{s=\{+,-\}}\!\!\!\!\mu_{s}N_{s}}{\ddagger}+\frac{2k_{\text{F}}a}{\pi\nu}N_{+}N_{-}\nonumber\\
 &\!\!\!\!-&\frac{8k_{\text{F}}^{2}a^{2}}{\pi^{2}\nu^{2}V^{3}}\sum_{\vec{k}_{1,2,3}}\!\frac{n_{+}(\epsilon_{\vec{k}_{1}})n_{-}(\epsilon_{\vec{k}_{2}})\left(n_{+}(\epsilon_{\vec{k}_{3}})+n_{-}(\epsilon_{\vec{k}_{3}})\right)}{\epsilon_{\vec{k}_{1}}+\epsilon_{\vec{k}_{2}}-\epsilon_{\vec{k}_{3}}-\epsilon_{\vec{k}_{4}}}\punc{.}
 \label{eqn:TheAFMFreeEnergy}
\end{eqnarray*}
This expression coincides
\footnote{The result of Ref.~\cite{05dm11} was a perturbation expansion to
second order in the scattering length $a$ considering all Green's function
contributions. The term labeled ($\ddagger$) corresponds to the ``$e^{(0)}-Ts$''
term of Ref.~\cite{05dm11} --- i.e. the difference between the kinetic energy
and entropy. The $\Delta\mu=0$ limit has also been derived
elsewhere~\cite{58ak05,96p07}.} with that obtained in Ref.~\cite{05dm11}. The
method employed in the numerical calculation of the summation over three momenta
is described in \appref{sec:ComputationalAnalysisOfMomentumSpaceIntegral}.

\subsection{Magnetization}\label{sec:magnetisationdirection}

To minimize the free energy and obtain the net magnetization it is convenient to
take the expression for the thermodynamic potential \eqref{eqn:popimbthermpot}
and affect the shift of the field
$\Phi_{\text{z}}\mapsto\Phi_{\text{z}}-\Delta\mu\pi\nu/2k_{\text{F}a}$. As a
result, the thermodynamic potential takes the form
\begin{eqnarray}
 &\Phi_{\text{G}}&=-\frac{1}{\beta V}\sum_{\blankfrac{\vec{k}}{s=\{+,-\}}}\ln\left(1+\e{-\beta(\epsilon_{\vec{k}}-\mu_{s})}\right)\nonumber\\
 &+&\!\!\frac{2k_{\text{F}}a}{\pi\nu}\left|{\bm\phi}_{0}-\frac{\Delta\mu\vec{e}_{\text{z}}\pi\nu}{2k_{\text{F}a}}\right|^{2}-\frac{2k_{\text{F}}a}{\pi\nu}\rho_{0}^{2}+\frac{2k_{\text{F}}a}{\pi\nu}N_{+}N_{-}\nonumber\\
 &-&\!\!\frac{8k_{\text{F}}^{2}a^{2}}{\pi^{2}\nu^{2}V^{3}}\sum_{\vec{k}_{1,2,3}}\frac{n_{+}(\epsilon_{\vec{k}_{1}})n_{-}(\epsilon_{\vec{k}_{2}})\left(n_{+}(\epsilon_{\vec{k}_{3}})+n_{-}(\epsilon_{\vec{k}_{3}})\right)}{\epsilon_{\vec{k}_{1}}+\epsilon_{\vec{k}_{2}}-\epsilon_{\vec{k}_{3}}-\epsilon_{\vec{k}_{4}}}\punc{,}\nonumber\\
\end{eqnarray}
where, in response to the shift of $\Phi_{\text{z}}$, the factors of
$\mu_{s}=\mu-2k_{\text{F}}a\rho_{0}/\pi\nu+2k_{\text{F}}as|{\bm\phi}_{0}|/\pi\nu$
entering the definitions of $N_\pm$ and $n_\pm$ are now independent of
$\Delta\mu$. The thermodynamic potential can be rewritten in terms of a function
of just the auxiliary fields and the chemical potential shift as
$\Phi_{\text{G}}=F(|{\bm\phi}_{0}|)+2k_{\text{F}}a|{\bm\phi}_{0}-
\Delta\mu\vec{e}_{\text{z}}\pi\nu/2k_{\text{F}a}|^{2}/\pi\nu-2k_{\text{F}}a\rho_{0}^{2}/\pi\nu$.

In the grand canonical ensemble, the thermodynamic potential must be minimized
with respect to the components of the auxiliary field giving
\begin{equation}
 \frac{F'(|{\bm\phi}_{0}|)}{|{\bm\phi}_{0}|}\left(\begin{array}{c}{\bm\phi}_{\perp}\\\phi_{\text{z}}\end{array}\right)+\frac{4k_{\text{F}}a}{\pi\nu}\left(\begin{array}{c}{\bm\phi}_{\perp}\\\phi_{\text{z}}-\Delta\mu\pi\nu/2k_{\text{F}a}\end{array}\right)=\vec{0}\punc{,}
\end{equation}
where ${\bm\phi}_{\perp}=(\phi_{\text{x}},\phi_{\text{y}})$ so
${\bm\phi}_{0}={\bm\phi}_{\perp}+\phi_{\text{z}}\vec{e}_{\text{z}}$. Following
\secref{sec:FerroOrderingAuxiliaryFieldMeanFieldSaddlePointValues} one may now
identify the magnetization with the field
${\bm\phi}_{0}-\pi\nu\Delta\mu\vec{e}_{\text{z}}/2k_{\text{F}}a$. If
$\Delta\mu=0$, then the system of equations is solved by either
$F'(|{\bm\phi}_{0}|)/|{\bm\phi}_{0}|+4k_{\text{F}}a/\pi\nu=0$ (the
direction of spontaneous ferromagnetism in-plane remains undetermined), or
${\bm\phi}_{0}=\vec{0}$. If $\Delta\mu\ne0$ then
${\bm\phi}_{\perp}=\vec{0}$, and the magnetization is set by the equation
$F'(\phi_{\text{z}})=2(\Delta\mu-2k_{\text{F}}a\phi_{\text{z}}/\pi\nu)$ and is
oriented along the z-axis. This behavior is analogous to what we saw in the
mean-field analysis in
\secref{sec:FerroOrderingAuxiliaryFieldMeanFieldSaddlePointValues}. Finally, as
a consistency check, one may note that the expected degree of population
imbalance can be recovered from the grand potential
$M=-\left.{\partial\Phi_{\text{G}}}/{\partial\Delta\mu}\right|_{T,V,N}$.

\section{Population imbalance}\label{sec:Results}

\begin{figure}
 \centerline{\resizebox{0.6\linewidth}{!}{\includegraphics{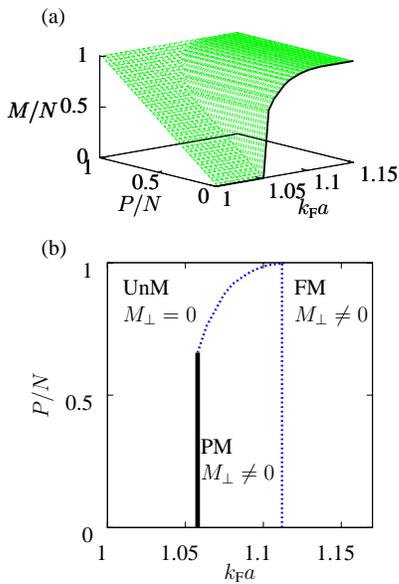}}}
 \caption{(a) shows the magnetization $M$ as a function of
   population imbalance, $P$ and interaction strength
   $k_{\text{F}}a=\sqrt[3]{3\pi^{2}(n_{\uparrow}+n_{\downarrow})}$ in the canonical
   ensemble at $T=0$ at fixed species populations. The thick line traces system
   variation at $P=0$ which corresponds to trap profile ($P/N=0$) in
   \figref{fig:trapprofall}. (b) shows the phase boundary between
   ``unmagnetized'' (UnM) and partially magnetized (PM) region and the line of
   saturation before the fully-magnetized (FM) region. Note that, by
   unmagnetized, we refer to the not in-plane magnetization. The solid line
   denotes first order transitions, the dashed second order and saturation.}
 \label{fig:PhaseTransCanonical}
\end{figure}

With the formal part of the analysis complete, we will now apply these results
to explore the implications of ferromagnetism in the atomic Fermi gas. To begin,
let us consider the phase behavior of the system in the canonical ensemble
working at fixed particle number. The variation of the total magnetization,
$|M|$, as a function of interaction strength and particle imbalance can be found
by minimizing the free energy at fixed particle number. The results are shown in
\figref{fig:PhaseTransCanonical}. To ensure that the free energy is locally
\emph{minimized} rather than just being at a \emph{stationary
value}~\cite{07sr04}, the curvature was examined numerically. In the balanced
Fermi gas, $P=0$, the results shown in \figref{fig:PhaseTransCanonical}(a)
recapitulate those discussed by \citet{05dm11}. In particular at zero
temperature, when the interaction strength is small,
$k_{\text{F}}a\lesssim1.05$, there is no net magnetization. As the interaction
strength is increased, at $k_{\text{F}}a\approx1.05$ there is a first order
phase transition into a magnetized phase with $M/N\approx0.6$. As
$k_{\text{F}}a$ is increased further the magnetization rises until it is
saturated at $k_{\text{F}}a\approx1.11$.

With increasing population imbalance, $P$, at $k_{\text{F}}a\lesssim1.05$, where
it is not energetically favorable for a spontaneous magnetization to develop,
the magnetization is forced to stay pinned to the minimum value set by the
imbalance. With increasing interaction strength, at $k_{\text{F}}a\approx1.05$
there is a first order transition and the magnetization jumps to
$M/N\approx0.6$. This feature is consistent with the
findings of the Stoner mean-field theory that the transition interaction
strength found is independent of population imbalance. If the population imbalance is
greater than $P/N\gtrsim0.6$ then the magnetization takes the value of the
spontaneous magnetization projected onto the sheet of minimum magnetization
caused by the population imbalance.

From these results, one can infer the corresponding zero temperature phase
diagram \figref{fig:PhaseTransCanonical}(b). Characterizing the phase behavior
by the strength of the in-plane magnetization and the degree of polarization, the
phase diagram divides into three distinct regions. At low interaction strength
the system is not spontaneously unmagnetized, though there can be a
magnetization fixed by the population imbalance. Then, at increased interaction
strength the system become partially magnetized either through a first order (at
low population imbalance) or a second order phase transition. At interaction
strength above $k_{\text{F}}a\gtrsim1.11$ the magnetization saturates.

To address the properties of the population imbalanced system in the grand
canonical regime, we will divide our discussion between the uniform and trap
geometries. In \secref{sec:TrappingPotentialUniformSystem} we will address the
properties of a uniform system where the chemical potential $\mu$ and shift
$\Delta\mu$ are held constant (allowing the species populations to effectively
interchange). Drawing on these results, we will then discuss the phase behavior
in a harmonic trap in \secref{sec:TrappingPotentialTrappedSystem}.

\subsection{Uniform system}\label{sec:TrappingPotentialUniformSystem}

\begin{figure}
 \centerline{\resizebox{\linewidth}{!}{\includegraphics{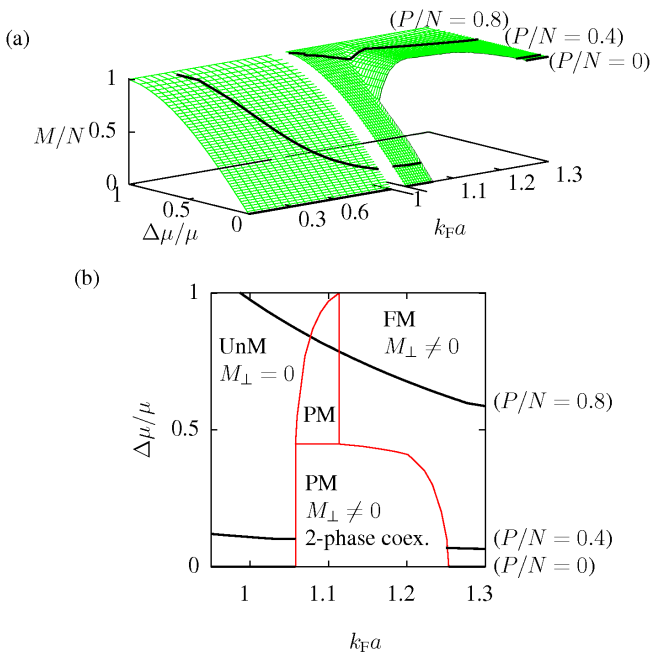}}}
 \caption{(Color online) (a) shows the variation of magnetization $M$ as a
   function of chemical potential shift $\Delta\mu$ and interaction strength
   $k_{\text{F}}a$ in the grand canonical ensemble at $T=0$. The thick lines
   correspond to trap profiles at $P/N=0$, $P/N=0.4$, and $P/N=0.8$ in
   \figref{fig:trapprofall}. In the region where magnetization is undefined
   there is phase separation. The lower set of diagrams show the phase
   boundaries (and saturation line) between ``unmagnetized'' (UnM), partially
   magnetized (PM) and fully-magnetized (FM) regions, as well as the region of
   phase separation.}
 \label{fig:PhaseTransGrand}
\end{figure}

In the spatially uniform system, when the chemical potentials of the two species
are fixed, for each value of the interaction strength $k_{\text{F}}a$ and
relative shift in chemical potential $\Delta\mu/\mu$, from the free energy one
can obtain the phase corresponding to minimal thermodynamic potential. Applying
this procedure, the resulting phase behavior is shown in
\figref{fig:PhaseTransGrand}. For $\Delta\mu/\mu=0$ and small interaction
strength $k_{\text{F}}a\lesssim1.05$ there is no magnetization. As the
interaction strength is increased, at $k_{\text{F}}a\approx1.05$ in the
canonical regime \figref{fig:PhaseTransCanonical} there is a first order phase
transition into a fully-magnetized state. Working at fixed chemical potential
[\figref{fig:PhaseTransGrand}(a)], the phase transition straight into a
saturated state increases the number of particles, which in turn increases the
effective interaction strength to $k_{\text{F}}a\approx1.25$ (calculated using
the chemical potential for a non-interacting system with the same total number
of particles). This leads to an intermediate region of phase separation in the
grand canonical regime. At
$k_{\text{F}}a\lesssim1.05$ as the chemical potential shift is increased up to
$\Delta\mu/\mu=1$, the magnetization increases up to its maximum saturated value
as the Fermi surfaces become more unbalanced. At $\Delta\mu/\mu>1$ the chemical
potential of the minority spin species is negative so only the majority spin
species remain and the system is fully magnetized. With a chemical potential
shift the region of phase separation corresponds to the first order phase
transition in \figref{fig:PhaseTransCanonical}. The corresponding phase diagram
showing the regime of two-phase coexistence is shown in
\figref{fig:PhaseTransGrand}(b).

Finally, if the system has an imposed density and population imbalance, and the
chemical potentials are free to vary, then there are two possibilities: Firstly,
the spontaneous ferromagnetism is sufficient to provide the population imbalance
and any excess magnetization lies in the plane. This corresponds to a point on
the line $\Delta\mu=0$ in \figref{fig:PhaseTransGrand}. The second
possibility is that spontaneous ferromagnetism is not sufficient, and so there
is an additional chemical potential shift $\Delta\mu\ne0$. In this case the
magnetization then points along the direction of population imbalance. This is
consistent with the findings in \secref{sec:magnetisationdirection}. For a given
interaction strength, the magnetization increases with chemical potential shift
to saturation, so there is always a chemical potential shift that will give a
suitable population imbalance.

\subsection{Trapped system}\label{sec:TrappingPotentialTrappedSystem}

Using the insight gained from the study of the uniform system, we can now
explore an atomic Fermi gas in the physical system --- a potential trap. Without
loss of generality we take $\uparrow$ ($\downarrow$) to represent the majority
(minority) species of atoms. We focus on a harmonic trap, with rescaled spatial
coordinates to ensure a spherically symmetric trapping potential,
$V(\vec{r})\sim r^{2}$. Furthermore, we make use of the local density
approximation in which the chemical potential of both species
$\mu_{\text{eff},\sigma}(\vec{r})=\mu_{0,\sigma}-V(\vec{r})$ are renormalized
by the same trapping potential. Although there is some experimental evidence
\cite{06plklh01,06pllhhs11} that the local density approximation might not be
valid \cite{06dm05,06ibld11} in some setups, we believe that its application
here will correctly address the qualitative phase structure. The chemical
potentials are regarded to be locally fixed, therefore the local phase is that
of the uniform system in the grand canonical regime examined in
\secref{sec:TrappingPotentialUniformSystem}. With a constant chemical potential
shift $\Delta\mu$ and interaction strength $g$, but varying effective chemical
potential $\mu$, the system follows the trajectory
$k_{\text{F}}a\propto\sqrt{\mu}$ and $\Delta\mu/\mu\propto1/\mu$ in the grand
canonical regime shown in \figref{fig:PhaseTransGrand}. If the chemical
potential is large, the system spontaneously becomes ferromagnetic, and the
magnetization is saturated; if the chemical potential is small, the relative
chemical potential shift is large ensuring the magnetization is again near
saturation. The locus in \figref{fig:PhaseTransGrand} shows that, in the
intermediate region, the magnetization can develop a minimum depending on the
degree of population imbalance.

\begin{figure}
 \centerline{\resizebox{0.7\linewidth}{!}{\includegraphics{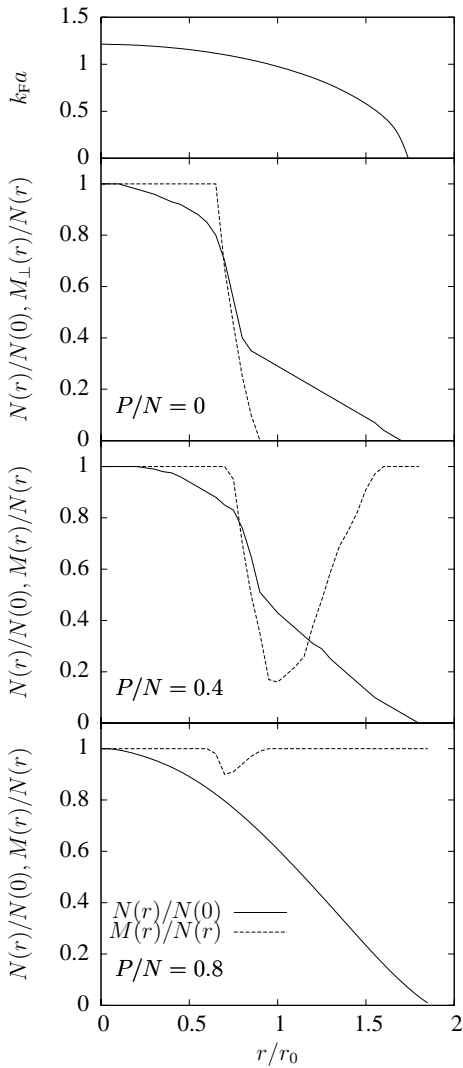}}}
 \caption{The density of particles at radius $r$ in a trap potential profile at
  three different values of total population imbalance. The variation in the
  local particle density $N$ is shown by the solid line, and the variation in
  the local magnetization $M$ is shown by the dashed line. The plot densities
  are renormalized by their central density, $N_{0}$, radii by the outer radius,
  $r_{0}$, of non-interacting particles with the same average inner chemical
  potential, $\mu_{0}$. The upper panel shows the effective $k_{\text{F}}a$ in
  the $P/N=0$ case.}
 \label{fig:trapprofall}
\end{figure}

To understand the behavior in the trap geometry, one should note the following:
If the degree of equilibrium pseudo-spin magnetization is in excess of that
imposed by total population imbalance alone, the analysis of
\secref{sec:magnetisationdirection} tells us that some component of the
spontaneous magnetization lies along the z-axis with the remainder oriented in
the x-y plane. If, however, net population imbalance is large, then
$\Delta\mu\ne0$ and no in-plane magnetization develops. Here one may identify
three characteristic behaviors with radial density profiles shown in
\figref{fig:trapprofall}. The first ($P/N=0$) has in-plane magnetization, and
the others do not. The second ($P/N=0.4$) has a first order transition and
non-zero phase separation whereas the third ($P/N=0.8$) is always fully
magnetized due to strong interactions. The three plots all have the same central
chemical potential.

The first possibility shown in \figref{fig:trapprofall}($P/N=0$) is at small
population imbalance, involving the development of a spontaneous magnetization
which is in excess of what can be absorbed by population imbalance alone, in
this case $\Delta\mu=0$ and some magnetization lies in the plane. At small
radii, where the interaction strength $k_{\text{F}}a>1.25$ is greater than the
limit for ferromagnetism, the results of the uniform system
(\secref{sec:TrappingPotentialUniformSystem}) show that there is saturated
ferromagnetism in the plane and a normal component that provides the fixed
population imbalance. Following this there is a region of phase separation and
then at $k_{\text{F}}a\approx1.05$ there are equal particle densities and no
magnetization. The outer edge of the particle distribution of both species is
where $\mu_{0}=V(r_{0})$.

In the second scenario shown in \figref{fig:trapprofall}($P/N=0.4$) the
spontaneous magnetization is not sufficient to provide population imbalance
alone, in this case we require $\Delta\mu\ne0$, and all magnetization is
oriented along the axis of population imbalance. From the trap center the
population imbalance is first fully saturated, followed by a region of phase
separation, into a region of partial magnetization.  This causes the minority
spin particles to have a sharp maximum number density at $r/r_{0}\approx0.6$,
and the magnetization to have a corresponding minimum; this counters the
intuitive expectation that number density should rise towards the trap center
due to the increasing effective chemical potential. As the effective chemical
potential continues to fall with increasing radius, the minority spin species
population falls more rapidly than the majority and magnetization increases. At
a large radius, the chemical potential of the minority spin particles reaches
zero before the majority spin so there is a thin shell containing only majority
spin particles at the outside and so is fully magnetized.

The third possibility shown in \figref{fig:trapprofall}($P/N=0.8$) is that the
locus in \figref{fig:PhaseTransGrand} does not cross the first order transition
and region of phase separation. At $\Delta\mu/\mu<1$ the system is fully
magnetized due to the strong interactions between particles. At
$\Delta\mu/\mu>1$ the system is fully magnetized due to there being no minority
spin particles. In the intermediate regime there is a narrow band where the
system is partially polarized. The majority spin species exists out to greater
radius than in cases ($P/N=0$) and ($P/N=0.4$) because $\Delta\mu$ is larger so
a greater potential at a larger radius is required to give the majority spin
species zero effective chemical potential.

\section{Discussion}

To conclude, let us now consider four methods of how spin magnetization could be
detected experimentally. Firstly, the interaction energy can be estimated by
studying the expansion of the gas \cite{03bckmkss07}. Time of flight measurements of
the expanding cloud with no external magnetic field $B=0$ are ballistic and so
can provide the initial kinetic energy. If the magnetic field is present,
$B\ne0$, then interactions are significant during the expansion.  Collisions ensure
that all of the interaction energy is converted into kinetic energy so the
measurements reflect the total released energy. Taking the difference between the
$B\ne0$ and $B=0$ measurements therefore probes the interaction energy. An
unmagnetized gas has interaction energy whereas the fully magnetized gas has zero
interaction energy so time of flight measurements should allow the ferromagnetic
state to be detected.

Radio frequency spectroscopy \cite{08kz01} allows one to probe the spatial
variations of scattering lengths by exciting the atoms from one spin state
$\ket{1}$ into some other state $\ket{3}$ whilst leaving the atoms in the second
spin state $\ket{2}$ unaffected. The presence of atoms in state $\ket{2}$ shifts
the resonance $\nu_{13}$ by $\Delta\nu_{13}=2n_{2}(a_{23}-a_{21})$, where
$a_{ij}$ is the scattering length between states $\ket{i}$ and
$\ket{j}$. Measurement of the resonance shift could allow the spatial
distribution of the individual species to be probed. The presence of the
ferromagnetic state could be inferred by looking for the characteristic density
profiles outlined in \secref{sec:TrappingPotentialUniformSystem}.

A third simple method of detecting a ferromagnetic transition could be to
monitor the size of the atomic cloud. In a harmonic trap the cloud size is
proportional to the square root of the Fermi energy. Therefore, the size of the
fully-magnetized  state is $2^{1/3}$ larger than the unmagnetized.

On the repulsive side of the Feshbach resonance three-body collisions can result
in the formation of a molecular bound state of two atoms that might destroy the
atomic gas before it has time to undergo ferromagnetic ordering. To overcome
this obstacle an atomic gas spin could be polarized along the magnetic field
direction and an RF $\pi/2$ pulse applied to rotate all the spins into the plane
\cite{08kz01}. The rate of precession of the spins is set by the magnetic field
strength, which varies across the atomic gas due to field inhomogeneities. The
precession rate of the atoms would however be kept locked together by the
ferromagnetic interaction. Furthermore the ferromagnetic phase has an
antisymmetric wave function which inhibits collisions and so prevents the
formation of molecular bound states. A signature of ferromagnetism is therefore
the absence of molecular bound state formation.

We now outline two possible ways to further our analysis. The first order phase
transition leads to discontinuities in the density and magnetization leading to
phase separation. Such behavior could lead to a breakdown of the local density
approximation, a potential source of inaccuracy in our analysis. This could
be fixed through inclusion of a surface energy.

The second is to investigate the possibility that magnetic texture could
develop. Textured modes may have been seen via the possible formation of a
CDW/SDW in experimental results on the analogous solid state systems of
itinerant electron ferromagnets $\text{UGe}_{2}$ \cite{01hsrkbcf03,02wm06},
$\text{Ca}_{3}\text{Ru}_{2}\text{O}_{7}$ \cite{06bikhmpsldrmhs03}, and MnSi
\cite{01pjl11}. Our general formalism should be able to be extended to include
the possibility of a textured phase which lies beyond the first order line in
the putative paramagnetic regime.

In conclusion we have developed a general formalism to describe itinerant
ferromagnetic transitions in two-component fermionic cold atom systems with
repulsive interactions, and potential population imbalance. At low population
imbalance, we predict that the first order transition that characterizes the
balanced system persists. However, when the imbalance is large the transition
becomes continuous. In the trap geometry we found the first order phase
transition led to discontinuities in density and magnetization. Up to a critical
total population imbalance, set by the possible total magnetization following a
first order transition, the phases in the trap had the same density and
magnetization profiles with increasing population imbalance, but in-plane
magnetization fell.  With population imbalance above this level, the system
requires a chemical potential shift to generate a population imbalance; however
there is still a small range over which a first order phase transition is
seen. In the two latter cases the local population imbalance displayed a
characteristic minimum with radius.

\acknowledgements The authors acknowledge the financial support of the EPSRC,
and thank Wolfgang Ketterle and Zoran Hadzibabic for useful discussions.

\appendix
\section{Computational analysis of momentum space integral}\label{sec:ComputationalAnalysisOfMomentumSpaceIntegral}

\begin{figure}
 \centerline{\resizebox{0.8\linewidth}{!}{\includegraphics{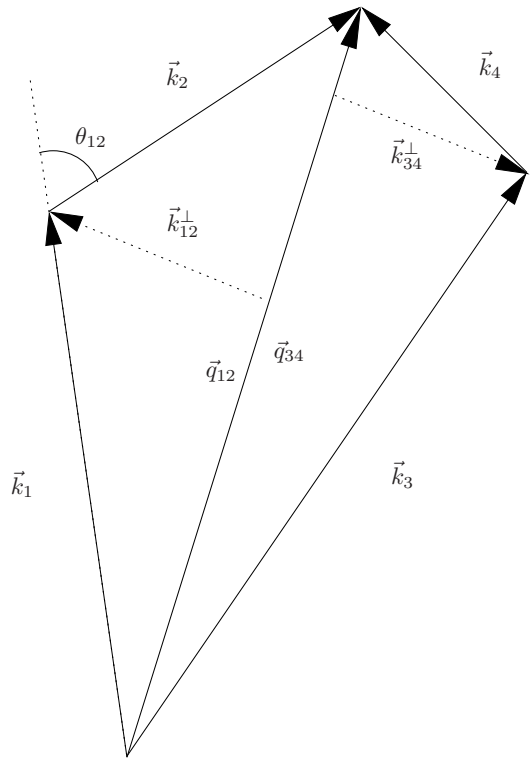}}}
 \caption{The re-parameterization of momenta used to ensure momentum
  conservation. $\vec{k}_{1,2,3,4}$ represent the momenta appearing in the
  original integral, whose separate sums are
  $\vec{q}_{12}=\vec{k}_{1}+\vec{k}_{2}$ and
  $\vec{q}_{34}=\vec{k}_{3}+\vec{k}_{4}$, which the Dirac delta function in
  \eqnref{eqn:GeneralQuadkIntegralToBeReParameterised} will ensure that
  $\vec{q}_{12}=\vec{q}_{34}$. $\theta_{12}$ represents the angle between
  $\vec{k}_{1}$ and $\vec{k}_{2}$, $\vec{k}_{12}^{\perp}$ is the vector
  perpendicular from $\vec{q}_{12}$ to $\vec{k}_{1}$ and $\vec{k}_{2}$, and
  $\vec{k}_{34}^{\perp}$ is similarly defined.}
 \label{fig:FourthOrderMomSpaceInt}
\end{figure}

An important integral \eqnref{eqn:TheAFMFreeEnergy} encountered in this paper
has the form
\begin{eqnarray}
 \label{eqn:GeneralQuadkIntegralToBeReParameterised}
 &&\iiiint F(|\vec{k}_{1}|,|\vec{k}_{2}|,|\vec{k}_{3}|,|\vec{k}_{4}|)\delta(\vec{k}_{1}+\vec{k}_{2}-\vec{k}_{3}-\vec{k}_{4})\nonumber\\
 &&\times d\vec{k}_{1}d\vec{k}_{2}d\vec{k}_{3}d\vec{k}_{4}\punc{.}
\end{eqnarray}
To evaluate this integral one could substitute
$\vec{k}_{4}=\vec{k}_{1}+\vec{k}_{2}-\vec{k}_{3}$, and then integrate over three
parameters representing the lengths of vectors $\vec{k}_{1}$, $\vec{k}_{2}$, and
$\vec{k}_{3}$, and a minimum of three relative angles between these vectors,
giving a total of six integration parameters. However, numerical integration
generally becomes more prohibitive with increasing number of dimensions. Since
the function $F$ depends only on the magnitude of the momentum, the scheme
outlined below allows us to perform the angular integration separately of the
function and leave a numerical integral over just the four dimensions of the
vector lengths.

The integral is re-parameterized according to
\figref{fig:FourthOrderMomSpaceInt}, $\vec{q}_{12}=\vec{k}_{1}+\vec{k}_{2}$ and
$\vec{q}_{34}=\vec{k}_{3}+\vec{k}_{4}$, the vector perpendicular from
$\vec{q}_{12}$ to $\vec{k}_{1}$ and $\vec{k}_{2}$ is $\vec{k}_{12}^{\perp}$,
and $\vec{k}_{34}^{\perp}$ is similarly defined. The vector $\vec{k}_{12}^{\perp}$
has length given by
\begin{equation}
 k_{12}^{\perp}=\frac{1}{2q}\sqrt{2q^{2}\left(k_{1}^{2}+k_{2}^{2}\right)-q^{4}-\left(k_{1}^{2}-k_{2}^{2}\right)^{2}}\punc{.}
\end{equation}
We first concentrate on calculating the angular component just of the integral
over $\vec{k}_{1}$ and $\vec{k}_{2}$, the angle between these vectors
is $\theta_{12}$. The phase space volume of the angular integral is
\begin{eqnarray}
 \sin\theta_{12}d\theta_{12}&\!\!=\!\!\!\!&\frac{k_{12}^{\perp}}{k_{1}k_{2}}\left(\!2\!+\!\sqrt{\frac{k_{1}^{2}-k_{12}^{\perp2}}{k_{2}^{2}-k_{12}^{\perp2}}}+\sqrt{\frac{k_{2}^{2}-k_{12}^{\perp2}}{k_{1}^{2}-k_{12}^{\perp2}}}\right)d k_{12}^{\perp}\nonumber\\
 &\!\!=\!\!&-\frac{q_{12}}{k_{1}k_{2}}d q_{12}\punc{,}
\end{eqnarray}
where $|k_{1}-k_{2}|\le q_{12}\le k_{1}+k_{2}$. The total number density
integrated over two momenta can then be found using
\begin{eqnarray}
 \label{eqn:ReParameterisedIntegralDoublekIntoq}
 &&\int_{0}^{\pi}4\pi k_{1}^{2}2\pi k_{2}^{2}\sin\theta_{12}d\theta_{12}\nonumber\\
 &&=\int_{|k_{1}-k_{2}|}^{k_{1}+k_{2}}4\pi k_{1}2\pi k_{2}q_{12}d q_{12}=4\pi k_{1}^{2}4\pi k_{2}^{2}\punc{,}
\end{eqnarray}
which is the expected result. A similar procedure is used to parameterize the
separate integral over the angular components of $\vec{k}_{3}$ and $\vec{k}_{4}$
into $\vec{q}_{34}$.

The original integral \eqnref{eqn:GeneralQuadkIntegralToBeReParameterised} is
now re-written in terms of the parameters $q_{12}$ and $q_{34}$ using
\eqnref{eqn:ReParameterisedIntegralDoublekIntoq}. Momentum conservation is
required by the presence of the Dirac delta function
$\delta(\vec{k}_{1}+\vec{k}_{2}-\vec{k}_{3}-\vec{k}_{4})=\delta(\vec{q}_{12}-\vec{q}_{34})$,
however the $q_{12}$ and $q_{34}$ parameters introduced are just scalar
quantities. The momentum conservation requirement is implemented by demanding
that the two scalar integration parameters are equal, which sets the two
integration parameters equal, $q_{12}=q_{34}=q$ so there is just one integral
over parameter $q$ remaining. However, this introduces an extra angular degree
of freedom (the angle between $\vec{q}_{12}$ and $\vec{q}_{34}$). In order to
compensate the integrand is divided by the extra phase space volume of the
angular integration, $4\pi q^{2}$. We then obtain
\begin{eqnarray}
 &&16\pi^{3}\iiiint F(k_{1},k_{2},k_{3},k_{4})k_{1}k_{2}k_{3}k_{4}\nonumber\\
 &&\times\max[0,\min(k_{1}+k_{2},k_{3}+k_{4})\nonumber\\
 &&-\max(|k_{1}-k_{2}|,|k_{3}-k_{4}|)]d k_{1}d k_{2}d k_{3}d k_{4}\punc{.}
\end{eqnarray}
\\
\\
This integral is better suited to computational evaluation since it is
four-dimensional [rather than the six-dimensional
\eqnref{eqn:GeneralQuadkIntegralToBeReParameterised}], and the term introduced
to compensate for the angular integral has a relatively simple form.
\\


\begin{thebibliography}{54}
\expandafter\ifx\csname natexlab\endcsname\relax\def\natexlab#1{#1}\fi
\expandafter\ifx\csname bibnamefont\endcsname\relax
  \def\bibnamefont#1{#1}\fi
\expandafter\ifx\csname bibfnamefont\endcsname\relax
  \def\bibfnamefont#1{#1}\fi
\expandafter\ifx\csname citenamefont\endcsname\relax
  \def\citenamefont#1{#1}\fi
\expandafter\ifx\csname url\endcsname\relax
  \def\url#1{\texttt{#1}}\fi
\expandafter\ifx\csname urlprefix\endcsname\relax\def\urlprefix{URL }\fi
\providecommand{\bibinfo}[2]{#2}
\providecommand{\eprint}[2][]{\url{#2}}

\bibitem[{\citenamefont{Stwalley}(1976)}]{76s12}
\bibinfo{author}{\bibfnamefont{W.~C.} \bibnamefont{Stwalley}},
  \bibinfo{journal}{Phys. Rev. Lett.} \textbf{\bibinfo{volume}{37}},
  \bibinfo{pages}{1628} (\bibinfo{year}{1976}).

\bibitem[{\citenamefont{Tiesinga et~al.}(1993)\citenamefont{Tiesinga, Verhaar,
  and Stoof}}]{93tvs05}
\bibinfo{author}{\bibfnamefont{E.}~\bibnamefont{Tiesinga}},
  \bibinfo{author}{\bibfnamefont{B.~J.} \bibnamefont{Verhaar}},
  \bibnamefont{and} \bibinfo{author}{\bibfnamefont{H.~T.~C.}
  \bibnamefont{Stoof}}, \bibinfo{journal}{Phys. Rev. A}
  \textbf{\bibinfo{volume}{47}}, \bibinfo{pages}{4114} (\bibinfo{year}{1993}).

\bibitem[{\citenamefont{Strecker et~al.}(2003)\citenamefont{Strecker,
  Partridge, and Hulet}}]{03sph08}
\bibinfo{author}{\bibfnamefont{K.~E.} \bibnamefont{Strecker}},
  \bibinfo{author}{\bibfnamefont{G.~B.} \bibnamefont{Partridge}},
  \bibnamefont{and} \bibinfo{author}{\bibfnamefont{R.~G.} \bibnamefont{Hulet}},
  \bibinfo{journal}{Phys. Rev. Lett.} \textbf{\bibinfo{volume}{91}},
  \bibinfo{pages}{080406} (\bibinfo{year}{2003}).

\bibitem[{\citenamefont{Gupta et~al.}(2003)\citenamefont{Gupta, Hadzibabic,
  Stan, Dieckmann, Schunck, van Kempen, Verhaar, and Ketterle}}]{03ghzsdskvk06}
\bibinfo{author}{\bibfnamefont{S.}~\bibnamefont{Gupta}},
  \bibinfo{author}{\bibfnamefont{M.~W.} \bibnamefont{Hadzibabic},
  \bibfnamefont{Z.~Zwierlein}}, \bibinfo{author}{\bibfnamefont{C.~A.}
  \bibnamefont{Stan}}, \bibinfo{author}{\bibfnamefont{.~K.}
  \bibnamefont{Dieckmann}}, \bibinfo{author}{\bibfnamefont{C.~H.}
  \bibnamefont{Schunck}}, \bibinfo{author}{\bibfnamefont{E.~G.~M.}
  \bibnamefont{van Kempen}}, \bibinfo{author}{\bibfnamefont{B.~J.}
  \bibnamefont{Verhaar}}, \bibnamefont{and}
  \bibinfo{author}{\bibfnamefont{W.}~\bibnamefont{Ketterle}},
  \bibinfo{journal}{Science} \textbf{\bibinfo{volume}{300}},
  \bibinfo{pages}{1723} (\bibinfo{year}{2003}).

\bibitem[{\citenamefont{Chin et~al.}(2004)\citenamefont{Chin, Bartenstein,
  Altmeyer, Riedl, Jochim, Hecker~Denschlag, and Grimm}}]{04cbarjhg08}
\bibinfo{author}{\bibfnamefont{C.}~\bibnamefont{Chin}},
  \bibinfo{author}{\bibfnamefont{M.}~\bibnamefont{Bartenstein}},
  \bibinfo{author}{\bibfnamefont{A.}~\bibnamefont{Altmeyer}},
  \bibinfo{author}{\bibfnamefont{S.}~\bibnamefont{Riedl}},
  \bibinfo{author}{\bibfnamefont{S.}~\bibnamefont{Jochim}},
  \bibinfo{author}{\bibfnamefont{J.}~\bibnamefont{Hecker~Denschlag}},
  \bibnamefont{and} \bibinfo{author}{\bibfnamefont{R.}~\bibnamefont{Grimm}},
  \bibinfo{journal}{Science} \textbf{\bibinfo{volume}{305}},
  \bibinfo{pages}{1128} (\bibinfo{year}{2004}).

\bibitem[{\citenamefont{Regal et~al.}(2004)\citenamefont{Regal, Greiner, and
  Jin}}]{04rgj01}
\bibinfo{author}{\bibfnamefont{C.~A.} \bibnamefont{Regal}},
  \bibinfo{author}{\bibfnamefont{M.}~\bibnamefont{Greiner}}, \bibnamefont{and}
  \bibinfo{author}{\bibfnamefont{D.~S.} \bibnamefont{Jin}},
  \bibinfo{journal}{Phys. Rev. Lett.} \textbf{\bibinfo{volume}{92}},
  \bibinfo{pages}{040403} (\bibinfo{year}{2004}).

\bibitem[{\citenamefont{Stoner}(1937)}]{57s04}
\bibinfo{author}{\bibfnamefont{E.~C.} \bibnamefont{Stoner}},
  \bibinfo{journal}{Proc. R. Soc. London} \textbf{\bibinfo{volume}{165}},
  \bibinfo{pages}{372} (\bibinfo{year}{1937}).

\bibitem[{\citenamefont{Wohlfarth and Rhodes}(1962)}]{62wr11}
\bibinfo{author}{\bibfnamefont{E.~P.} \bibnamefont{Wohlfarth}}
  \bibnamefont{and} \bibinfo{author}{\bibfnamefont{P.}~\bibnamefont{Rhodes}},
  \bibinfo{journal}{Philos. Mag.} \textbf{\bibinfo{volume}{7}},
  \bibinfo{pages}{1817} (\bibinfo{year}{1962}).

\bibitem[{\citenamefont{Ziman}(1979)}]{79z11}
\bibinfo{author}{\bibfnamefont{J.~M.} \bibnamefont{Ziman}},
  \emph{\bibinfo{title}{Principles of the Theory of Solids}}
  (\bibinfo{publisher}{Cambridge University Press}, \bibinfo{year}{1979}).

\bibitem[{\citenamefont{Shimizu}(1964)}]{64s09}
\bibinfo{author}{\bibfnamefont{M.}~\bibnamefont{Shimizu}},
  \bibinfo{journal}{Proc. Phys. Soc. London} \textbf{\bibinfo{volume}{84}},
  \bibinfo{pages}{397} (\bibinfo{year}{1964}).

\bibitem[{\citenamefont{Belitz et~al.}(1997)\citenamefont{Belitz, Kirkpatrick,
  and Vojta}}]{97bkv04}
\bibinfo{author}{\bibfnamefont{D.}~\bibnamefont{Belitz}},
  \bibinfo{author}{\bibfnamefont{T.~R.} \bibnamefont{Kirkpatrick}},
  \bibnamefont{and} \bibinfo{author}{\bibfnamefont{T.}~\bibnamefont{Vojta}},
  \bibinfo{journal}{Phys. Rev. B} \textbf{\bibinfo{volume}{55}},
  \bibinfo{pages}{9452} (\bibinfo{year}{1997}).

\bibitem[{\citenamefont{Belitz et~al.}(1999)\citenamefont{Belitz, Kirkpatrick,
  and Vojta}}]{99bkv06}
\bibinfo{author}{\bibfnamefont{D.}~\bibnamefont{Belitz}},
  \bibinfo{author}{\bibfnamefont{T.~R.} \bibnamefont{Kirkpatrick}},
  \bibnamefont{and} \bibinfo{author}{\bibfnamefont{T.}~\bibnamefont{Vojta}},
  \bibinfo{journal}{Phys. Rev. Lett.} \textbf{\bibinfo{volume}{82}},
  \bibinfo{pages}{4707} (\bibinfo{year}{1999}).

\bibitem[{\citenamefont{Vojta}(2000)}]{00v06}
\bibinfo{author}{\bibfnamefont{T.}~\bibnamefont{Vojta}}, \bibinfo{journal}{Ann.
  Phys.} \textbf{\bibinfo{volume}{9}}, \bibinfo{pages}{403}
  (\bibinfo{year}{2000}).

\bibitem[{\citenamefont{Belitz and Kirkpatrick}(2002)}]{02bk12}
\bibinfo{author}{\bibfnamefont{D.}~\bibnamefont{Belitz}} \bibnamefont{and}
  \bibinfo{author}{\bibfnamefont{T.~R.} \bibnamefont{Kirkpatrick}},
  \bibinfo{journal}{Phys. Rev. Lett.} \textbf{\bibinfo{volume}{89}},
  \bibinfo{pages}{247202} (\bibinfo{year}{2002}).

\bibitem[{\citenamefont{Belitz et~al.}(2005)\citenamefont{Belitz, Kirkpatrick,
  and Rollb\"uhler}}]{05bkr06}
\bibinfo{author}{\bibfnamefont{D.}~\bibnamefont{Belitz}},
  \bibinfo{author}{\bibfnamefont{T.~R.} \bibnamefont{Kirkpatrick}},
  \bibnamefont{and}
  \bibinfo{author}{\bibfnamefont{J.}~\bibnamefont{Rollb\"uhler}},
  \bibinfo{journal}{Phys. Rev. Lett.} \textbf{\bibinfo{volume}{94}},
  \bibinfo{pages}{247205} (\bibinfo{year}{2005}).

\bibitem[{\citenamefont{Uhlarz et~al.}(2004)\citenamefont{Uhlarz, Pfleiderer,
  and Hayden}}]{04uph12}
\bibinfo{author}{\bibfnamefont{M.}~\bibnamefont{Uhlarz}},
  \bibinfo{author}{\bibfnamefont{C.}~\bibnamefont{Pfleiderer}},
  \bibnamefont{and} \bibinfo{author}{\bibfnamefont{S.~M.}
  \bibnamefont{Hayden}}, \bibinfo{journal}{Phys. Rev. Lett.}
  \textbf{\bibinfo{volume}{93}}, \bibinfo{pages}{256404}
  (\bibinfo{year}{2004}).

\bibitem[{\citenamefont{Uhlarz et~al.}(2005)\citenamefont{Uhlarz, Pfleiderer,
  and Hayden}}]{05uph04}
\bibinfo{author}{\bibfnamefont{M.}~\bibnamefont{Uhlarz}},
  \bibinfo{author}{\bibfnamefont{C.}~\bibnamefont{Pfleiderer}},
  \bibnamefont{and} \bibinfo{author}{\bibfnamefont{S.~M.}
  \bibnamefont{Hayden}}, \bibinfo{journal}{Physica B}
  \textbf{\bibinfo{volume}{359-361}}, \bibinfo{pages}{1174}
  (\bibinfo{year}{2005}).

\bibitem[{\citenamefont{Huxley et~al.}(2000)\citenamefont{Huxley, Sheikin, and
  Braithwaite}}]{00hsb07}
\bibinfo{author}{\bibfnamefont{A.}~\bibnamefont{Huxley}},
  \bibinfo{author}{\bibfnamefont{I.}~\bibnamefont{Sheikin}}, \bibnamefont{and}
  \bibinfo{author}{\bibfnamefont{D.}~\bibnamefont{Braithwaite}},
  \bibinfo{journal}{Physica B} \textbf{\bibinfo{volume}{284-288}},
  \bibinfo{pages}{1277} (\bibinfo{year}{2000}).

\bibitem[{\citenamefont{Pfleiderer et~al.}(2001)\citenamefont{Pfleiderer,
  Julian, and Lonzarich}}]{01pjl11}
\bibinfo{author}{\bibfnamefont{C.}~\bibnamefont{Pfleiderer}},
  \bibinfo{author}{\bibfnamefont{S.~R.} \bibnamefont{Julian}},
  \bibnamefont{and} \bibinfo{author}{\bibfnamefont{G.~G.}
  \bibnamefont{Lonzarich}}, \bibinfo{journal}{Nature (London)}
  \textbf{\bibinfo{volume}{414}}, \bibinfo{pages}{427} (\bibinfo{year}{2001}).

\bibitem[{\citenamefont{Yu et~al.}(2004)\citenamefont{Yu, Zamborszky, Thompson,
  Sarrao, Torelli, Fisk, and Brown}}]{04yztstfb02}
\bibinfo{author}{\bibfnamefont{W.}~\bibnamefont{Yu}},
  \bibinfo{author}{\bibfnamefont{F.}~\bibnamefont{Zamborszky}},
  \bibinfo{author}{\bibfnamefont{J.~D.} \bibnamefont{Thompson}},
  \bibinfo{author}{\bibfnamefont{J.~L.} \bibnamefont{Sarrao}},
  \bibinfo{author}{\bibfnamefont{M.~E.} \bibnamefont{Torelli}},
  \bibinfo{author}{\bibfnamefont{Z.}~\bibnamefont{Fisk}}, \bibnamefont{and}
  \bibinfo{author}{\bibfnamefont{S.~E.} \bibnamefont{Brown}},
  \bibinfo{journal}{Phys. Rev. Lett.} \textbf{\bibinfo{volume}{92}},
  \bibinfo{pages}{086403} (\bibinfo{year}{2004}).

\bibitem[{\citenamefont{Uemura et~al.}(2007)\citenamefont{Uemura, Goko,
  Gat-Malureanu, Carlo, Russo, Savici, Acze, MacDougall, Rodriguez, Luke
  et~al.}}]{07uggcrsamrldmapbybflss01}
\bibinfo{author}{\bibfnamefont{Y.}~\bibnamefont{Uemura}},
  \bibinfo{author}{\bibfnamefont{T.}~\bibnamefont{Goko}},
  \bibinfo{author}{\bibfnamefont{I.}~\bibnamefont{Gat-Malureanu}},
  \bibinfo{author}{\bibfnamefont{J.}~\bibnamefont{Carlo}},
  \bibinfo{author}{\bibfnamefont{P.}~\bibnamefont{Russo}},
  \bibinfo{author}{\bibfnamefont{A.}~\bibnamefont{Savici}},
  \bibinfo{author}{\bibfnamefont{A.}~\bibnamefont{Acze}},
  \bibinfo{author}{\bibfnamefont{G.}~\bibnamefont{MacDougall}},
  \bibinfo{author}{\bibfnamefont{J.}~\bibnamefont{Rodriguez}},
  \bibinfo{author}{\bibfnamefont{G.}~\bibnamefont{Luke}}, \bibnamefont{et~al.},
  \bibinfo{journal}{Nature (London)} \textbf{\bibinfo{volume}{3}},
  \bibinfo{pages}{29} (\bibinfo{year}{2007}).

\bibitem[{\citenamefont{Otero-Leal et~al.}(2009)\citenamefont{Otero-Leal,
  Rivadulla, Saxena, Ahilan, and Rivas}}]{09orsar02}
\bibinfo{author}{\bibfnamefont{M.}~\bibnamefont{Otero-Leal}},
  \bibinfo{author}{\bibfnamefont{F.}~\bibnamefont{Rivadulla}},
  \bibinfo{author}{\bibfnamefont{S.~S.} \bibnamefont{Saxena}},
  \bibinfo{author}{\bibfnamefont{K.}~\bibnamefont{Ahilan}}, \bibnamefont{and}
  \bibinfo{author}{\bibfnamefont{J.}~\bibnamefont{Rivas}},
  \bibinfo{journal}{Phys. Rev. B} \textbf{\bibinfo{volume}{79}},
  \bibinfo{pages}{060401(R)} (\bibinfo{year}{2009}).

\bibitem[{\citenamefont{Petrova et~al.}(2009)\citenamefont{Petrova,
  Krasnorussky, Lograsso, and Stishov}}]{09pkls03}
\bibinfo{author}{\bibfnamefont{A.~E.} \bibnamefont{Petrova}},
  \bibinfo{author}{\bibfnamefont{V.~N.} \bibnamefont{Krasnorussky}},
  \bibinfo{author}{\bibfnamefont{T.~A.} \bibnamefont{Lograsso}},
  \bibnamefont{and} \bibinfo{author}{\bibfnamefont{S.~M.}
  \bibnamefont{Stishov}}, \bibinfo{journal}{Phys. Rev. B}
  \textbf{\bibinfo{volume}{79}}, \bibinfo{pages}{100401(R)}
  (\bibinfo{year}{2009}).

\bibitem[{\citenamefont{Otero-Leal et~al.}(2008)\citenamefont{Otero-Leal,
  Rivadulla, Garcia-Hernandez, Pineiro, Pardo, Baldomir, and
  Rivas}}]{08orgppbr06}
\bibinfo{author}{\bibfnamefont{M.}~\bibnamefont{Otero-Leal}},
  \bibinfo{author}{\bibfnamefont{F.}~\bibnamefont{Rivadulla}},
  \bibinfo{author}{\bibfnamefont{M.}~\bibnamefont{Garcia-Hernandez}},
  \bibinfo{author}{\bibfnamefont{A.}~\bibnamefont{Pineiro}},
  \bibinfo{author}{\bibfnamefont{V.}~\bibnamefont{Pardo}},
  \bibinfo{author}{\bibfnamefont{D.}~\bibnamefont{Baldomir}}, \bibnamefont{and}
  \bibinfo{author}{\bibfnamefont{J.}~\bibnamefont{Rivas}},
  \textbf{\bibinfo{volume}{arXiv:cond-mat/0806.2819v1 [cond-mat.str-el]}}
  (\bibinfo{year}{2008}).

\bibitem[{\citenamefont{Misawa et~al.}(2008)\citenamefont{Misawa, Yamaji, and
  Imada}}]{08myi09}
\bibinfo{author}{\bibfnamefont{T.}~\bibnamefont{Misawa}},
  \bibinfo{author}{\bibfnamefont{Y.}~\bibnamefont{Yamaji}}, \bibnamefont{and}
  \bibinfo{author}{\bibfnamefont{M.}~\bibnamefont{Imada}}, \bibinfo{journal}{J.
  Phys. Jpn.} \textbf{\bibinfo{volume}{77}}, \bibinfo{pages}{093712}
  (\bibinfo{year}{2008}).

\bibitem[{\citenamefont{Hamlin et~al.}(2007)\citenamefont{Hamlin, Deemyad,
  Schilling, Jacobsen, Kumar, Cornelius, Cao, and Neumeier}}]{07hdsjkccn07}
\bibinfo{author}{\bibfnamefont{J.~J.} \bibnamefont{Hamlin}},
  \bibinfo{author}{\bibfnamefont{S.}~\bibnamefont{Deemyad}},
  \bibinfo{author}{\bibfnamefont{J.~S.} \bibnamefont{Schilling}},
  \bibinfo{author}{\bibfnamefont{M.~K.} \bibnamefont{Jacobsen}},
  \bibinfo{author}{\bibfnamefont{R.~S.} \bibnamefont{Kumar}},
  \bibinfo{author}{\bibfnamefont{A.~L.} \bibnamefont{Cornelius}},
  \bibinfo{author}{\bibfnamefont{G.}~\bibnamefont{Cao}}, \bibnamefont{and}
  \bibinfo{author}{\bibfnamefont{J.~J.} \bibnamefont{Neumeier}},
  \bibinfo{journal}{Phys. Rev. B} \textbf{\bibinfo{volume}{76}},
  \bibinfo{pages}{014432} (\bibinfo{year}{2007}).

\bibitem[{\citenamefont{Liu and Hui}(2006)}]{06lh07}
\bibinfo{author}{\bibfnamefont{X.-J.} \bibnamefont{Liu}} \bibnamefont{and}
  \bibinfo{author}{\bibfnamefont{H.}~\bibnamefont{Hui}},
  \bibinfo{journal}{Europhys. Lett.} \textbf{\bibinfo{volume}{75}},
  \bibinfo{pages}{364} (\bibinfo{year}{2006}).

\bibitem[{\citenamefont{Combescot}(2007)}]{07c02}
\bibinfo{author}{\bibfnamefont{R.}~\bibnamefont{Combescot}}, in
  \emph{\bibinfo{booktitle}{Ultra-cold Fermi Gases}}, edited by
  \bibinfo{editor}{\bibfnamefont{M.}~\bibnamefont{Inguscio}},
  \bibinfo{editor}{\bibfnamefont{W.}~\bibnamefont{Ketterle}}, \bibnamefont{and}
  \bibinfo{editor}{\bibfnamefont{C.}~\bibnamefont{Salomon}}
  (\bibinfo{publisher}{IOS Press}, \bibinfo{year}{2007}), vol.
  \bibinfo{volume}{164}, p. \bibinfo{pages}{697}.

\bibitem[{\citenamefont{Sheehy and Radzihovsky}(2007{\natexlab{a}})}]{07sr08}
\bibinfo{author}{\bibfnamefont{D.~E.} \bibnamefont{Sheehy}} \bibnamefont{and}
  \bibinfo{author}{\bibfnamefont{L.}~\bibnamefont{Radzihovsky}},
  \bibinfo{journal}{Ann. Phys.} \textbf{\bibinfo{volume}{322}},
  \bibinfo{pages}{1790} (\bibinfo{year}{2007}{\natexlab{a}}).

\bibitem[{\citenamefont{Sogo and Yabu}(2002)}]{02sy10}
\bibinfo{author}{\bibfnamefont{T.}~\bibnamefont{Sogo}} \bibnamefont{and}
  \bibinfo{author}{\bibfnamefont{H.}~\bibnamefont{Yabu}},
  \bibinfo{journal}{Phys. Rev. A} \textbf{\bibinfo{volume}{66}},
  \bibinfo{pages}{043611} (\bibinfo{year}{2002}).

\bibitem[{\citenamefont{Duine and MacDonald}(2005)}]{05dm11}
\bibinfo{author}{\bibfnamefont{R.~A.} \bibnamefont{Duine}} \bibnamefont{and}
  \bibinfo{author}{\bibfnamefont{A.~H.} \bibnamefont{MacDonald}},
  \bibinfo{journal}{Phys. Rev. Lett.} \textbf{\bibinfo{volume}{95}},
  \bibinfo{pages}{230403} (\bibinfo{year}{2005}).

\bibitem[{\citenamefont{Zhang et~al.}(2008)\citenamefont{Zhang, Hung, and
  Wu}}]{08zhw05}
\bibinfo{author}{\bibfnamefont{S.}~\bibnamefont{Zhang}},
  \bibinfo{author}{\bibfnamefont{H.-h.} \bibnamefont{Hung}}, \bibnamefont{and}
  \bibinfo{author}{\bibfnamefont{C.}~\bibnamefont{Wu}},
  \textbf{\bibinfo{volume}{arXiv:cond-mat/0805.3031v5 [cond-mat.str-el]}}
  (\bibinfo{year}{2008}).

\bibitem[{\citenamefont{Huxley et~al.}(2001)\citenamefont{Huxley, Sheikin,
  Ressouche, Kernavanois, Braithwaite, Calemczuk, and Flouquet}}]{01hsrkbcf03}
\bibinfo{author}{\bibfnamefont{A.}~\bibnamefont{Huxley}},
  \bibinfo{author}{\bibfnamefont{I.}~\bibnamefont{Sheikin}},
  \bibinfo{author}{\bibfnamefont{E.}~\bibnamefont{Ressouche}},
  \bibinfo{author}{\bibfnamefont{N.}~\bibnamefont{Kernavanois}},
  \bibinfo{author}{\bibfnamefont{D.}~\bibnamefont{Braithwaite}},
  \bibinfo{author}{\bibfnamefont{R.}~\bibnamefont{Calemczuk}},
  \bibnamefont{and} \bibinfo{author}{\bibfnamefont{J.}~\bibnamefont{Flouquet}},
  \bibinfo{journal}{Phys. Rev. B} \textbf{\bibinfo{volume}{63}},
  \bibinfo{pages}{144519} (\bibinfo{year}{2001}).

\bibitem[{\citenamefont{Watanabe and Miyake}(2002)}]{02wm06}
\bibinfo{author}{\bibfnamefont{S.}~\bibnamefont{Watanabe}} \bibnamefont{and}
  \bibinfo{author}{\bibfnamefont{K.}~\bibnamefont{Miyake}},
  \bibinfo{journal}{J. Phys. and Chem. Solids} \textbf{\bibinfo{volume}{63}},
  \bibinfo{pages}{1465} (\bibinfo{year}{2002}).

\bibitem[{\citenamefont{Kitagawa et~al.}(2005)\citenamefont{Kitagawa, Ishida,
  Perry, Tayama, Sakakibara, and Maeno}}]{05kiptsm09}
\bibinfo{author}{\bibfnamefont{K.}~\bibnamefont{Kitagawa}},
  \bibinfo{author}{\bibfnamefont{K.}~\bibnamefont{Ishida}},
  \bibinfo{author}{\bibfnamefont{R.~S.} \bibnamefont{Perry}},
  \bibinfo{author}{\bibfnamefont{T.}~\bibnamefont{Tayama}},
  \bibinfo{author}{\bibfnamefont{T.}~\bibnamefont{Sakakibara}},
  \bibnamefont{and} \bibinfo{author}{\bibfnamefont{Y.}~\bibnamefont{Maeno}},
  \bibinfo{journal}{Phys. Rev. Lett.} \textbf{\bibinfo{volume}{95}},
  \bibinfo{pages}{127001} (\bibinfo{year}{2005}).

\bibitem[{\citenamefont{Berdnikov et~al.}(2008)\citenamefont{Berdnikov,
  Coleman, and Simon}}]{08bcs05}
\bibinfo{author}{\bibfnamefont{I.}~\bibnamefont{Berdnikov}},
  \bibinfo{author}{\bibfnamefont{P.}~\bibnamefont{Coleman}}, \bibnamefont{and}
  \bibinfo{author}{\bibfnamefont{S.~H.} \bibnamefont{Simon}},
  \textbf{\bibinfo{volume}{arXiv:0805.3693v1 [cond-mat.str-el]}}
  (\bibinfo{year}{2008}).

\bibitem[{\citenamefont{Keiter}(1970)}]{70k11}
\bibinfo{author}{\bibfnamefont{H.}~\bibnamefont{Keiter}},
  \bibinfo{journal}{Phys. Rev. B} \textbf{\bibinfo{volume}{2}},
  \bibinfo{pages}{3777} (\bibinfo{year}{1970}).

\bibitem[{\citenamefont{Evenson et~al.}(1970)\citenamefont{Evenson, Schrieffer,
  and Wang}}]{70esw03}
\bibinfo{author}{\bibfnamefont{W.~E.} \bibnamefont{Evenson}},
  \bibinfo{author}{\bibfnamefont{J.~R.} \bibnamefont{Schrieffer}},
  \bibnamefont{and} \bibinfo{author}{\bibfnamefont{S.~Q.} \bibnamefont{Wang}},
  \bibinfo{journal}{J. Appl. Phys.} \textbf{\bibinfo{volume}{41}},
  \bibinfo{pages}{1199} (\bibinfo{year}{1970}).

\bibitem[{\citenamefont{Morandi et~al.}(1974)\citenamefont{Morandi,
  Galleani~d'Agliano, Napoli, and Ratto}}]{74mgnr11}
\bibinfo{author}{\bibfnamefont{G.}~\bibnamefont{Morandi}},
  \bibinfo{author}{\bibfnamefont{E.}~\bibnamefont{Galleani~d'Agliano}},
  \bibinfo{author}{\bibfnamefont{F.}~\bibnamefont{Napoli}}, \bibnamefont{and}
  \bibinfo{author}{\bibfnamefont{C.~F.} \bibnamefont{Ratto}},
  \bibinfo{journal}{Adv. Phys.} \textbf{\bibinfo{volume}{23}},
  \bibinfo{pages}{867} (\bibinfo{year}{1974}).

\bibitem[{\citenamefont{Hubbard}(1979)}]{79h03}
\bibinfo{author}{\bibfnamefont{J.}~\bibnamefont{Hubbard}},
  \bibinfo{journal}{Phys. Rev. B} \textbf{\bibinfo{volume}{19}},
  \bibinfo{pages}{2626} (\bibinfo{year}{1979}).

\bibitem[{\citenamefont{Prange and Korenman}(1979{\natexlab{a}})}]{79pk05IV}
\bibinfo{author}{\bibfnamefont{R.~E.} \bibnamefont{Prange}} \bibnamefont{and}
  \bibinfo{author}{\bibfnamefont{V.}~\bibnamefont{Korenman}},
  \bibinfo{journal}{Phys. Rev. B} \textbf{\bibinfo{volume}{19}},
  \bibinfo{pages}{4691} (\bibinfo{year}{1979}{\natexlab{a}}).

\bibitem[{\citenamefont{Kanamori}(1963)}]{63k09}
\bibinfo{author}{\bibfnamefont{J.}~\bibnamefont{Kanamori}},
  \bibinfo{journal}{Prog. Theor. Phys.} \textbf{\bibinfo{volume}{30}},
  \bibinfo{pages}{275} (\bibinfo{year}{1963}).

\bibitem[{\citenamefont{Hubbard}(1963)}]{63h11}
\bibinfo{author}{\bibfnamefont{J.}~\bibnamefont{Hubbard}},
  \bibinfo{journal}{Proc. R. Soc. London} \textbf{\bibinfo{volume}{276}},
  \bibinfo{pages}{238} (\bibinfo{year}{1963}).

\bibitem[{\citenamefont{Abrikosov and Khalatnikov}(1958)}]{58ak05}
\bibinfo{author}{\bibfnamefont{A.~A.} \bibnamefont{Abrikosov}}
  \bibnamefont{and} \bibinfo{author}{\bibfnamefont{I.~M.}
  \bibnamefont{Khalatnikov}}, \bibinfo{journal}{Soviet Phys. JETP}
  \textbf{\bibinfo{volume}{6}}, \bibinfo{pages}{888} (\bibinfo{year}{1958}).

\bibitem[{\citenamefont{Prange and Korenman}(1979{\natexlab{b}})}]{79pk05V}
\bibinfo{author}{\bibfnamefont{R.~E.} \bibnamefont{Prange}} \bibnamefont{and}
  \bibinfo{author}{\bibfnamefont{V.}~\bibnamefont{Korenman}},
  \bibinfo{journal}{Phys. Rev. B} \textbf{\bibinfo{volume}{19}},
  \bibinfo{pages}{4698} (\bibinfo{year}{1979}{\natexlab{b}}).

\bibitem[{\citenamefont{Pathria}(1996)}]{96p07}
\bibinfo{author}{\bibfnamefont{R.~K.} \bibnamefont{Pathria}},
  \emph{\bibinfo{title}{Statistical Mechanics}}
  (\bibinfo{publisher}{Butterworths, London}, \bibinfo{year}{1996}).

\bibitem[{\citenamefont{Sheehy and Radzihovsky}(2007{\natexlab{b}})}]{07sr04}
\bibinfo{author}{\bibfnamefont{D.~E.} \bibnamefont{Sheehy}} \bibnamefont{and}
  \bibinfo{author}{\bibfnamefont{L.}~\bibnamefont{Radzihovsky}},
  \bibinfo{journal}{Phys. Rev. B} \textbf{\bibinfo{volume}{75}},
  \bibinfo{pages}{136501} (\bibinfo{year}{2007}{\natexlab{b}}).

\bibitem[{\citenamefont{Partridge
  et~al.}(2006{\natexlab{a}})\citenamefont{Partridge, Li, Kamar, Liao, and
  Hulet}}]{06plklh01}
\bibinfo{author}{\bibfnamefont{G.~B.} \bibnamefont{Partridge}},
  \bibinfo{author}{\bibfnamefont{W.}~\bibnamefont{Li}},
  \bibinfo{author}{\bibfnamefont{R.~I.} \bibnamefont{Kamar}},
  \bibinfo{author}{\bibfnamefont{Y.}~\bibnamefont{Liao}}, \bibnamefont{and}
  \bibinfo{author}{\bibfnamefont{R.~G.} \bibnamefont{Hulet}},
  \bibinfo{journal}{Science} \textbf{\bibinfo{volume}{311}},
  \bibinfo{pages}{503} (\bibinfo{year}{2006}{\natexlab{a}}).

\bibitem[{\citenamefont{Partridge
  et~al.}(2006{\natexlab{b}})\citenamefont{Partridge, Li, Liao, Hulet, Haque,
  and Stoof}}]{06pllhhs11}
\bibinfo{author}{\bibfnamefont{G.~B.} \bibnamefont{Partridge}},
  \bibinfo{author}{\bibfnamefont{W.}~\bibnamefont{Li}},
  \bibinfo{author}{\bibfnamefont{Y.~A.} \bibnamefont{Liao}},
  \bibinfo{author}{\bibfnamefont{R.~G.} \bibnamefont{Hulet}},
  \bibinfo{author}{\bibfnamefont{M.}~\bibnamefont{Haque}}, \bibnamefont{and}
  \bibinfo{author}{\bibfnamefont{H.~T.~C.} \bibnamefont{Stoof}},
  \bibinfo{journal}{Phys. Rev. Lett.} \textbf{\bibinfo{volume}{97}},
  \bibinfo{pages}{190407} (\bibinfo{year}{2006}{\natexlab{b}}).

\bibitem[{\citenamefont{De~Silva and Mueller}(2006)}]{06dm05}
\bibinfo{author}{\bibfnamefont{T.~N.} \bibnamefont{De~Silva}} \bibnamefont{and}
  \bibinfo{author}{\bibfnamefont{E.~J.} \bibnamefont{Mueller}},
  \bibinfo{journal}{Phys. Rev. A} \textbf{\bibinfo{volume}{73}},
  \bibinfo{pages}{051602(R)} (\bibinfo{year}{2006}).

\bibitem[{\citenamefont{Imambekov et~al.}(2006)\citenamefont{Imambekov, Bolech,
  Lukin, and Demler}}]{06ibld11}
\bibinfo{author}{\bibfnamefont{A.}~\bibnamefont{Imambekov}},
  \bibinfo{author}{\bibfnamefont{C.~J.} \bibnamefont{Bolech}},
  \bibinfo{author}{\bibfnamefont{M.}~\bibnamefont{Lukin}}, \bibnamefont{and}
  \bibinfo{author}{\bibfnamefont{E.}~\bibnamefont{Demler}},
  \bibinfo{journal}{Phys. Rev. A} \textbf{\bibinfo{volume}{74}},
  \bibinfo{pages}{053626} (\bibinfo{year}{2006}).

\bibitem[{\citenamefont{Bourdel et~al.}(2003)\citenamefont{Bourdel, Cubizolles,
  Khaykovich, Magalhaes, Kokkelmans, Shlyapnikov, and Salomon}}]{03bckmkss07}
\bibinfo{author}{\bibfnamefont{T.}~\bibnamefont{Bourdel}},
  \bibinfo{author}{\bibfnamefont{J.}~\bibnamefont{Cubizolles}},
  \bibinfo{author}{\bibfnamefont{L.}~\bibnamefont{Khaykovich}},
  \bibinfo{author}{\bibfnamefont{K.~M.~F.} \bibnamefont{Magalhaes}},
  \bibinfo{author}{\bibfnamefont{S.~J.~J.~M.~F.} \bibnamefont{Kokkelmans}},
  \bibinfo{author}{\bibfnamefont{G.~V.} \bibnamefont{Shlyapnikov}},
  \bibnamefont{and} \bibinfo{author}{\bibfnamefont{C.}~\bibnamefont{Salomon}},
  \bibinfo{journal}{Phys. Rev. Lett.} \textbf{\bibinfo{volume}{91}},
  \bibinfo{pages}{020402} (\bibinfo{year}{2003}).

\bibitem[{\citenamefont{Ketterle and Zwierlein}(2008)}]{08kz01}
\bibinfo{author}{\bibfnamefont{W.}~\bibnamefont{Ketterle}} \bibnamefont{and}
  \bibinfo{author}{\bibfnamefont{M.~W.} \bibnamefont{Zwierlein}},
  \textbf{\bibinfo{volume}{arXiv:cond-mat/0801.2500v1 [cond-mat.other]}}
  (\bibinfo{year}{2008}).

\bibitem[{\citenamefont{Baumberger et~al.}(2006)\citenamefont{Baumberger,
  Ingle, Kikugawa, Hossain, Meevasana, Perry, Shen, Lu, Damascelli, Rost
  et~al.}}]{06bikhmpsldrmhs03}
\bibinfo{author}{\bibfnamefont{F.}~\bibnamefont{Baumberger}},
  \bibinfo{author}{\bibfnamefont{N.~J.~C.} \bibnamefont{Ingle}},
  \bibinfo{author}{\bibfnamefont{N.}~\bibnamefont{Kikugawa}},
  \bibinfo{author}{\bibfnamefont{M.~A.} \bibnamefont{Hossain}},
  \bibinfo{author}{\bibfnamefont{W.}~\bibnamefont{Meevasana}},
  \bibinfo{author}{\bibfnamefont{R.~S.} \bibnamefont{Perry}},
  \bibinfo{author}{\bibfnamefont{K.~M.} \bibnamefont{Shen}},
  \bibinfo{author}{\bibfnamefont{D.~H.} \bibnamefont{Lu}},
  \bibinfo{author}{\bibfnamefont{A.}~\bibnamefont{Damascelli}},
  \bibinfo{author}{\bibfnamefont{A.}~\bibnamefont{Rost}}, \bibnamefont{et~al.},
  \bibinfo{journal}{Phys. Rev. Lett.} \textbf{\bibinfo{volume}{96}},
  \bibinfo{pages}{107601} (\bibinfo{year}{2006}).

\end{thebibliography}

\end{document}